\documentclass[letterpaper]{article}
\usepackage{nlkgq}
\nocopyright
\usepackage[hyphens]{url}
\usepackage{graphicx}
\usepackage{natbib}
\usepackage{caption}
\usepackage{booktabs}
\usepackage{multirow}
\usepackage{listings}
\title{Natural Language Knowledge Graph Query Execution:\\Leveraging Controlled Semantics in the LLM Context Window}

\author{
    Blake G. Fitch
}
\affiliations{
    Max Planck Institute for Biological Cybernetics, T\"ubingen, Germany\\
    blake.fitch@tuebingen.mpg.de
}

\begin{document}
\maketitle

\begin{abstract}
Large Language Model (LLM) applications often transfer domain concepts into the model's context informally, through prompt prose, schema dumps, and examples. We show that for database queries, data model concepts pass to LLMs more effectively through representations whose vocabulary terms carry declared, machine-readable semantics (\emph{controlled semantics}). NLKGQ is a working system and reusable framework that does this for data modeled in a knowledge graph. A formal OWL ontology serves as the transfer mechanism, concentrating the meaning of the data into semantically precise tokens the model can use directly. In a single LLM call, NLKGQ places in the context a system prompt instructing on SPARQL, the complete domain OWL ontology, and a domain-specific prompt addition, together with the user's natural language query. The model then generates the SPARQL query directly, zero-shot. Where the native vocabulary of an existing database or federation of databases is opaque, a wrapper ontology substitutes clean terms and a runtime rewriter restores the native forms. Evaluating on DBLP-QuAD~2.0 showed that its scores depend on the graph snapshot, the endpoint used, and the wording of its machine-generated questions, so we propose DBLP-QuAD~3.1, which maintains the intent of 2.0 while making reference results deterministic, revising reference SPARQL where needed, and rewriting the natural language questions, with a frontier model, to state each reference query's intent clearly and completely. We evaluate on the DBLP-QuAD~2.0 benchmark (57.6\% Match under deterministic re-scoring), DBLP-QuAD~3.1 (89.9\% Match on 1,000 questions), SemOpenAlex (98\% Match against a published baseline's 86\% on the identical test set), and neuroimaging metadata (100\%).\end{abstract}

\section{Introduction}

Natural language query execution on knowledge graphs requires the LLM to know the data model: what classes exist, how they connect, what types their properties carry. Prior work transfers this knowledge informally (schema dumps, few-shot examples, prompt instructions) and then compensates with engineering complexity: fine-tuning~\citep{pan2025firesparql,mecharnia2025finetuned}, multi-call agentic exploration~\citep{grasp2025,dobriy2026agentic}, or retrieval-augmented pipelines~\citep{smeros2025sparqlllm}. A consistent finding across this literature is that providing schema information in context matters more than any of these design choices~\citep{gashkov2025memorization,emonet2024sparql,doulaverakis2025medical}.

This paper shows that careful design of the schema information in the context window, together with the runtime support, delivers more accuracy than published fine-tuned, agentic, and retrieval systems report. Formal OWL ontologies transfer data model concepts as semantically precise tokens that map directly to correct query patterns. Informal descriptions (column names, natural language instructions, schema comments) transfer ambiguous tokens the LLM must interpret probabilistically. Complete declared semantics for the vocabulary in the context window avoid leaving the model to invent domain concepts and relationships. We call this \emph{controlled semantics}: schema tokens in the context window carry declared, machine-readable meaning. The original NLKGQ paper introduced the system for natural language knowledge graph querying and established this on a single domain: the representation of the ontology, not the model or the prompt, determined accuracy~\citep{fitch2026nlkgq}. Here we test the principle on public knowledge graphs whose vocabularies we do not control, and on larger benchmarks.

The NLKGQ execution mechanism is a single LLM call. This OpenAI-compatible API call~\citep{openai_api} comprises a user prompt carrying the user's natural language query and a system prompt containing a generic set of instructions for SPARQL generation, the complete domain OWL ontology, and a domain-specific set of instructions (the \emph{domain rider}). The model generates the SPARQL query zero-shot. The system strips any non-SPARQL text (code fences, reasoning traces, commentary) from the response before execution. Where the native vocabulary of an existing database or federation is opaque, a \textbf{wrapper ontology} substitutes clean, formally typed terms and a deterministic runtime rewriter restores the native forms before endpoint execution. The native vocabulary of DBLP, the computer science bibliography's knowledge graph~\citep{ackermann2024dblpkg}, illustrates the problem: \texttt{dblp:publishedIn} yields a venue name as a plain string while \texttt{dblp:publishedInStream} yields the venue as a URI, and nothing in either name says so. Our \textbf{\texttt{dblpx} wrapper ontology} declares \texttt{dblpx:venueName} with range \texttt{xsd:string} and \texttt{dblpx:venue} with range \texttt{Venue}, so the distinction the model needs is well defined.

This insight originates in our original NLKGQ results on a purpose-built domain: a neuroimaging metadata KG whose ontology was designed end-to-end for natural language querying reached 100\% accuracy on 21 competency questions, against 43\% for the same questions over a SQL schema auto-generated from that ontology, rising only to 57\% with the ontology's annotations carried into column comments. This paper extends the approach to three domains and establishes three contributions:

\begin{enumerate}
\item \textbf{A demonstration that controlled semantics can deliver high accuracy without fine-tuning, agentic exploration, or very large models.} On the SemOpenAlex~\citep{farber2023semopenalex} benchmark the architecture scores 98/100 against the published baseline's 86/100 on the same test set. The DBLP wrapper ontology adds +25.3 exact-match points on 1,000 DBLP questions. The same single-call architecture reaches 89.9\% Match on DBLP (Qwen3.6-27B). A ten-model sweep across four families (Qwen, Gemma, Mistral, Llama) shows the two strongest models converging near 90\%. On neuroimaging metadata it reaches 100\%.

\item \textbf{Wrapper ontologies and a runtime rewriter.} These bring controlled semantics to databases and federations whose vocabularies we do not control: the LLM sees clean, typed terms, and a deterministic rewriter restores the native forms for execution on an unchanged endpoint such as the public DBLP KG~\citep{ackermann2024dblpkg}.

\item \textbf{A refined benchmark.} We identify systematic quality issues in DBLP-QuAD~2.0~\citep{taffa2025dblpquad2} and propose DBLP-QuAD~3.1 (hereafter QuAD~2.0 and QuAD~3.1) with deterministic evaluation, revised reference queries, and questions that state each reference query's intent.
\end{enumerate}

\section{Related Work}

\textbf{Schema injection for SPARQL generation.}
\citet{emonet2024sparql} show that GPT-4o without schema context scores 0.08 F1 on bioinformatics SPARQL; with retrieved schema and example context plus query validation it reaches 0.91. Medical SPARQL generation reaches 11/11 with KG structure extracted from an example entry vs.\ 4/11 without~\citep{doulaverakis2025medical}. \citet{gashkov2025memorization} demonstrate that LLMs memorize popular KG schemas: models reproduce real Wikidata URIs even when URIs are masked in the prompt, so scores on widely used public KGs partly measure memorization rather than generalization. This makes less-memorized KGs such as DBLP a stricter test of generalization, and it implies that on unfamiliar vocabularies the context window is the model's only source of schema knowledge. These results establish that schema information matters, but \emph{which formal properties} of the schema matter, and how to optimize them for LLM consumption, remains largely unexamined.

\textbf{SPARQL generation systems.}
GRASP~\citep{grasp2025} uses ReAct-style runtime exploration. Its feedback variant achieves 51.0\% F1 on 50 samples of the original DBLP-QuAD~\citep{banerjee2023dblpquad} with GPT-4.1 across multiple LLM calls. Its error analysis notes failures on citation queries where the OMID (OpenCitations Meta Identifier) indirection chain is too complex to discover at runtime, a case the \texttt{dblpx} wrapper handles by stating the relationship upfront. GRISP~\citep{grisp2026} fine-tunes skeleton generation with runtime IRI retrieval, stating that ``zero-shot agentic methods take the upper hand'' in low-data regimes. FIRESPARQL~\citep{pan2025firesparql} fine-tunes LLaMA-8B for scholarly KGs, achieving 85\% on SciQA but 0\% zero-shot. SPARQL-LLM~\citep{smeros2025sparqlllm} validates retrieval-augmented generation with schema and example injection plus a revision loop, and in doing so finds 11 defective reference queries in the challenge set it evaluated on, precedent that benchmark auditing is part of sound evaluation. Our approach is single-call and zero-shot, relying entirely on formal ontology vocabulary rather than examples or exploration.

\textbf{Benchmarks.}
QuAD~2.0~\citep{taffa2025dblpquad2} is a large log-derived scholarly benchmark but has systematic quality issues we document below. Its authors' own few-shot system with entity linking reports 57.74\% F1 on it~\citep{taffa2025dblpquad2}. High F1 there requires reproducing each reference's arbitrary 10-row slice, which few-shot prompting in the benchmark's own query style encourages. A system returning the full answer set is penalized for every extra row (see the critique below). Our QuAD~3.1 revision is likely to improve all systems' results. For SemOpenAlex, Bartels et al.~\citeyearpar{bartels2025sparqltrans} release a 100-question test set with reference queries. Their best baseline, a 123B model prompted few-shot with entity mappings, scores 86/100. We evaluate on the same test set.

\textbf{Schema representation and query generation.}
\citet{sequeda2023benchmark} show SPARQL outperforms SQL 3$\times$ overall, with SQL falling to zero on the hardest schema quadrants, suggesting that the richer semantics of RDF/OWL schemas benefit query generation. The original NLKGQ paper tightened this comparison by deriving both backends from the same ontology with the same data and models: SPARQL 100\%, SQL 57\%~\citep{fitch2026nlkgq}. \citet{mecharnia2025finetuned} find that fine-tuning works on KGs with readable names (DBpedia, 61\%) but fails on opaque identifiers (Wikidata, 13\%): direct evidence that vocabulary readability determines accuracy independently of model capability. The same principle is surfacing in the relational world: \citet{zhang2026sqlfriendly} propose semantics-preserving schema renaming and view-based abstraction to make SQL schemas legible to LLMs. Wrapper ontologies share a lineage with ontology-based data access~\citep{xiao2018obda}, which maps a clean conceptual vocabulary onto native schemas. Ours differs in being designed for LLM consumption following the NLKGQ design principles, and applied by a deterministic SPARQL-to-SPARQL rewrite to the native vocabulary before submission.

\section{Formal Ontologies as Concept Transfer}

NLKGQ generates SPARQL in a single LLM call. The prompt contains a system prompt with SPARQL rules, the full ontology in Turtle, an optional domain rider, and the user's question. Architectural details are in the original NLKGQ paper~\citep{fitch2026nlkgq}. Here we focus on \emph{why} the ontology representation matters. Pretraining has exposed LLMs to SPARQL and to the principles of querying knowledge graphs. Models write competent SPARQL over generic concepts without assistance. What a model lacks for a specific graph is the domain vocabulary: which classes and properties exist, how they connect, and what types they carry. The ontology in the context window supplies exactly this, so query generation reduces to concept transfer.

\subsection{What OWL Provides That SQL Does Not}

An OWL ontology in context gives the LLM five categories of formally grounded tokens that informal schema descriptions lack:

\begin{enumerate}
\item \textbf{Explicit domain and range.} Every property declares what class it connects and what type it expects. The LLM knows that \texttt{dblpx:signatureAuthor} links a \texttt{Signature} to a \texttt{Person} without inferring from data.

\item \textbf{Class hierarchy.} \texttt{rdfs:subClassOf} relationships let the LLM reason about generalization. A query about ``publications'' can include \texttt{Article}, \texttt{Inproceedings}, and \texttt{Book} without enumeration.

\item \textbf{Property typing.} \texttt{owl:ObjectProperty} vs.\ \texttt{owl:DatatypeProperty} tells the LLM whether to expect a URI join or a literal filter. SQL has foreign keys, but the semantics are implicit.

\item \textbf{Human-readable annotations.} \texttt{rdfs:label} and \texttt{rdfs:comment} attach natural language meaning to formal terms, directly in the vocabulary the LLM processes.

\item \textbf{Semantic naming.} Property names like \texttt{hasAuthor}, \texttt{publishedInStream}, \texttt{yearOfPublication} encode directional relationships in readable English. SQL column names (\texttt{auth\_id}, \texttt{pub\_yr}) are opaque.
\end{enumerate}

\subsection{Prior Results on a Controlled Domain}

The original NLKGQ paper introduced the system on neuroimaging metadata with a purpose-built ontology~\citep{fitch2026nlkgq}. We summarize the findings this paper builds on. With the full OWL ontology in context, SPARQL generation reached 100\% on 21 competency questions; auto-generated SQL against the same data reached 57\%. Stripping \texttt{rdfs:comment} annotations dropped SPARQL from 100\% to 81\%; reducing the ontology to bare graph structure dropped it to 5\%. Same data, model, and prompt; only the formal richness of the vocabulary varied. Three system-prompt variants tied once the vocabulary was right, and 35B MoE variants peaked at 57\% against the dense 27B's 100\%. That work also established six ontology design principles (full English words, descriptive property names, consistent \texttt{has}-prefixed naming, recognizable class names, explicit domain and range, no opaque identifiers); here we apply them to vocabularies we do not control.

\subsection{Wrapper Ontologies}

When a KG's native OWL vocabulary is itself problematic (opaque identifiers, namespace collisions, overloaded terms, missing annotations), we introduce a \emph{wrapper ontology}: a clean namespace with readable, unambiguous terms that we deterministically map to the native vocabulary. A SPARQL rewriter translates generated queries to native predicates and patterns before execution. One wrapper term may expand to several triples. The wrapper is an architectural feature of NLKGQ. This paper reports on two wrapper implementations, \texttt{dblpx} for DBLP and \texttt{oax} for SemOpenAlex.

For DBLP, the native vocabulary uses \texttt{dblp:Signature} for authorship reification, a name that suggests cryptography rather than authorship, has 19 external namespaces, and lacks \texttt{rdfs:comment}. The \texttt{dblpx} wrapper provides self-documenting property names (\texttt{signatureAuthor}, \texttt{signaturePosition}), explicit domain/range for every property, and a single namespace. The \texttt{dblpx} wrapper also spans a federation: our DBLP index loads the DBLP and OpenCitations~\citep{peroni2020opencitations} dumps together, and the single \texttt{dblpx} vocabulary covers entities and citations from both sources. For SemOpenAlex, the \texttt{oax} wrapper flattens indirect relationships (the Authorship/OpenAccess/Geo intermediary classes) into direct properties on the queried classes.

\subsection{A Worked Example}

The following case from the QuAD~3.1 benchmark runs shows the mechanism end to end. The question: \emph{``What are the author signatures for publications in venue \textless https://dblp.org/streams/conf/pods\textgreater\ from 2024? Return signature URI and publication URI, sorted by publication.''} The model, seeing only the \texttt{dblpx} wrapper vocabulary, generates (PREFIX declarations omitted):

\begin{lstlisting}
SELECT DISTINCT ?signature ?pub WHERE {
  ?pub a dblpx:Publication ;
    dblpx:venue <https://dblp.org/streams/conf/pods> ;
    dblpx:year "2024"^^xsd:gYear ;
    dblpx:hasSignature ?signature .
} ORDER BY ?pub
\end{lstlisting}

The rewriter deterministically restores the native vocabulary before execution:

\begin{lstlisting}
SELECT DISTINCT ?signature ?pub WHERE {
  ?pub a dblp:Publication ;
    dblp:publishedInStream
      <https://dblp.org/streams/conf/pods> ;
    dblp:yearOfPublication "2024"^^xsd:gYear ;
    dblp:hasSignature ?signature .
} ORDER BY ?pub
\end{lstlisting}

The model never sees \texttt{publishedInStream} or \texttt{yearOfPublication}. It works in a vocabulary where every term was chosen for semantic precision, and the mapping back to the endpoint's terms is mechanical.

\subsection{The Role of the Domain Rider}

NLKGQ's prompt contains natural language alongside the formal ontology: a system prompt establishing SPARQL discipline (the \emph{procedural} prompt cited in the table captions) and a short domain rider. With the ontology they form the \emph{concept layer}, the context tokens that carry out concept transfer. Our best configurations use both. The practical division of labor: domain knowledge that can be stated as vocabulary belongs in the ontology, where it accumulates safely. The rider is reserved for a small, stable set of query conventions. In our experiments the rider failed as a channel for incremental concept repair; the Discussion shows why.

\section{Benchmark Critique: DBLP-QuAD 2.0}

QuAD~2.0~\citep{taffa2025dblpquad2} comprises 5,000 question-SPARQL pairs derived from SPARQL query logs over DBLP. We evaluate against its 1,000-question test set and identify systematic issues that affect all systems evaluated on it.

\textbf{Non-deterministic evaluation.}
998 of 1,000 reference queries impose a \texttt{LIMIT} (992 of them \texttt{LIMIT~10}), 723 without \texttt{ORDER~BY}. Semantically equivalent queries with different triple pattern orderings return different row slices from the same engine. Under the benchmark's own protocol, re-executing the references at evaluation time, the share of cases our system matches varies run to run (52.1--52.9\% across five runs) as the reference slices change; with reference results computed once against the evaluation snapshot, the same system scores a stable 57.6\% (the QuAD~2.1 protocol). Either way, SPINACH~F1~\citep{liu2024spinach} stays near 17.5\%, while on our updated QuAD~3.1 (deterministic ordering) the same system scores 79.0\%. We attribute most of the 61.5-point F1 gap to the non-deterministic references. Match remains informative under this defect while row-wise F1 does not: the reference rows are an arbitrary slice of the true answer, and a correct generated query returns the full answer set, which contains whichever slice the engine produced for the reference. F1 counts every valid row outside that slice as a precision error, scoring correct queries down for the reference's arbitrariness. GRASP's error analysis notes unfairly low F1 from \texttt{LIMIT} differences~\citep{grasp2025}.

\textbf{LLM-generated questions.}
Natural language questions are generated by LLaMA-3.1-8B from SPARQL, producing unnatural phrasings and ambiguous intent where multiple SPARQL forms exist.

\textbf{Ambiguous entity references.}
Questions name entities in prose while the reference query resolves them to a specific URI. With many DBLP authors sharing identical names, the intended referent is often unrecoverable from the question text alone: systems with or without entity linking cannot reliably reproduce the reference result.

\textbf{Defective reference queries.}
122 reference queries are trivial echoes of the form \texttt{SELECT * WHERE \{ VALUES ?x \{ <uri> \} \}}: they return the URI or string from the question unchanged, testing nothing. Others use the wrong predicate for the stated intent (for example \texttt{dblp:createdBy}, which includes editors, where the question asks about authorship), omit \texttt{GROUP~BY} under aggregation, or carry malformed projections.

\textbf{Misdocumented data dependencies.}
211 cases require OpenCitations citation data. The benchmark's instructions point to the plain DBLP dump (\texttt{dblp.org/rdf/}), which contains no citation triples. The required combined DBLP+OpenCitations dump exists at \texttt{sparql.dblp.org} but is not referenced~\citep{ackermann2024dblpkg}. A system following the documentation cannot answer these cases. The statistics of \citet{ackermann2024dblpkg} imply that only 71\% of DBLP publications carry an OpenCitations identifier, so citation questions have an implicit completeness dependency no system can overcome.

\textbf{Reproducibility.}
The endpoint software is unspecified, the dump is not archived, and question generation ran on a closed institutional service.

\section{DBLP-QuAD 3.1}

We update QuAD~2.0 rather than build a new benchmark because its assets are worth preserving. It is a scholarly KGQA benchmark derived from real SPARQL query logs, and published systems have reported against it. As published, however, its scores conflate reference artifacts with system quality: the same system scores 17.5\% SPINACH~F1 on 2.0 and 79.0\% on 3.1, a gap we attribute mainly to the references, not the system. An unreleased QuAD~2.1 measurement protocol isolates the cause by keeping questions and reference queries unchanged. Reference results are computed once against the evaluation snapshot instead of re-executed on every run. Under it, Match rises from 52.9\% to 57.6\% and Exact from 11.3\% to 20.1\%, while SPINACH~F1 does not move (17.5\% to 17.3\%). We report 57.6\% as our QuAD~2.0 result. Consistent reference results repair individual cases, but the F1 metrics still score answer rows against an arbitrary slice of the reference result; only the deterministic references of 3.1 recover them (79.0\%). We therefore propose DBLP-QuAD~3.1, updating QuAD~2.0's 1,000 test cases while preserving their provenance. Every modification is logged per case. The repairs were made against the reference queries' intent, case by case, not against NLKGQ's outputs. The wrapper's advantage predates them (+9.9 Match on unmodified 2.0, Table~\ref{tab:progression} native vs.\ wrapper). The corrections:

\textbf{Deterministic evaluation.} Non-deterministic \texttt{LIMIT} clauses were removed (954 cases: 699 with no \texttt{ORDER~BY} and 255 where ties on the ordering key at the \texttt{LIMIT} boundary make the slice unstable), and under-constrained queries were scoped with year, venue, or affiliation filters so the natural result set is bounded (305 cases: 296 among the 954, 9 among the 44). \texttt{LIMIT} survives in 44 ranked top-$k$ queries whose \texttt{ORDER~BY} key is deterministic. Of these, 24 lacked \texttt{ORDER~BY} in 2.0 and received one matching the question's ranking intent (699 + 24 = 723). Reference results are now unique for semantically equivalent queries, up to ties on the ordering key in ranked cases.

\textbf{Reference query repair.} Corrected predicates, missing \texttt{GROUP~BY}, and broken aggregations, verified by execution against the documented endpoint. The 122 \texttt{VALUES}-only echo queries were replaced with real lookups on the referenced entity, preserving each question's intent.

\textbf{Higher-quality natural language.} Each machine-generated question was rewritten for natural phrasing and unambiguous intent using a frontier model (Claude), replacing the 8B-model paraphrases of QuAD~2.0. The 8B paraphrases were systematically more ambiguous, admitting multiple valid query forms; a per-case pass then amended 547 questions that omitted columns, ordering, or scope present in the reference.

\textbf{Explicit entity references.} 266 rewritten questions include bracketed entity URIs that the originals lacked. This is a deliberate scope decision: QuAD~3.1 evaluates query generation, not entity resolution. Where the original question named an entity ambiguously (common with author names), the URI pins the intended referent so that reference results are well defined. Systems with entity linking can ignore the brackets; systems without can still be evaluated on query construction. The convention is not benchmark-only: NLKGQ's web interface accepts the same bracketed URIs, giving users a direct way to pin the intended entity, and an entity linker could supply them automatically. We report this openly so 3.1 scores are not read as directly comparable to 2.0.

\textbf{Documented data dependencies.} The combined DBLP+OpenCitations dump requirement is stated explicitly, along with the 71\% citation coverage bound. The 28 federated cases, whose reference queries issue \texttt{SERVICE} calls to live public endpoints, are kept for continuity with QuAD~2.0. We note that their reference results depend on the availability and state of third-party SPARQL endpoints.

\textbf{Release.} QuAD~3.1 will be released on GitHub as a self-contained benchmark, independent of NLKGQ: the test cases together with scripts that recreate the evaluation snapshot of the combined DBLP+OpenCitations graph from persistent archives (DBLP's 2025-07-02 RDF release and the DBLP subset of the OpenCitations Index) and stand up a QLever instance on which its reference queries execute as published.

\section{Evaluation}

All experiments: single LLM call, temperature 0.0, no fine-tuning, no few-shot examples, no entity linking. SPARQL queries are executed on QLever~\citep{bast2017qlever}. All DBLP results in this paper run against the single fixed snapshot described above. The results are reproducible: models are open-weight and served locally on AMD APU nodes, generation at temperature 0.0 is near-deterministic, and repeated full-benchmark runs reproduced scores within 0.1 points on every deterministic configuration; only QuAD~2.0 with re-executed references varies more (the 52.1--52.9\% range above).

Table~\ref{tab:metrics} defines the four metrics we report. Given deterministic reference results, Match and Exact are strict and intuitive: a case either delivers the reference answer or it does not, which is what a user of the system experiences. We report the two F1 measures for comparability with published work, not as quality measures: both award partial credit when generated and reference results partially overlap, even when the question was not answered, and deduct for correct results that include extra rows or columns. QuAD F1 pools all result cells into one set per side; SPINACH F1 matches row to row. All reported values are means over the full case set, with failed cases (no executable query, or query error) scoring zero. To our knowledge, no published system on these benchmarks reports stricter than result containment.

\begin{table}[t]
\centering
\caption{Evaluation metrics, reported in this order.}
\label{tab:metrics}
\footnotesize
\begin{tabular}{@{}llp{4.6cm}@{}}
\toprule
Metric & Scoring & Criterion \\
\midrule
Match & binary & generated results contain the reference results, equal or superset \\
Exact & binary & generated results have an exact match to reference results \\
QuAD F1 & graded & set-of-values F1 over all result cells, flattened as in QuAD~2.0's evaluation code~\citep{taffa2025dblpquad2} \\
SPINACH F1 & graded & row-assignment F1 of \citet{liu2024spinach}, implemented following the published code of GRASP~\citep{grasp2025} \\
\bottomrule
\end{tabular}
\end{table}

\subsection{From QuAD 2.0 to QuAD 3.1}

Table~\ref{tab:progression} traces the path from the benchmark as published to our final configuration. The top row is the reproduction condition: QuAD~2.0 as released, with DBLP's native ontology.

Adding the \texttt{dblpx} wrapper (2.0-wrapper row) raises Match and Exact but \emph{lowers} both F1 measures. The better system scores worse. With non-deterministic references, a correct query returning the full answer set is penalized row for row against an arbitrary 10-row slice, and the \texttt{dblpx} wrapper's fuller, better-formed results widen that penalty. This inversion is the first evidence that on QuAD~2.0 the F1 columns measure the references, not the system.

The 2.1 row is QuAD~2.0 measured consistently: identical questions and reference queries, with reference results computed once against the evaluation snapshot rather than re-executed per run. (2.1 is a measurement protocol over the unchanged 2.0 artifact.) Match and Exact rise, and this row is our reported QuAD~2.0 result (57.6\% Match). Neither F1 column improves (QuAD~F1 17.6 to 16.7, SPINACH 17.5 to 17.3): consistent reference results repair some cases outright, yet the F1 metrics still score answer rows against an arbitrary slice of the reference result.

QuAD~3.1 (the two 3.1 rows) replaces the non-deterministic references with deterministic ones, repairs the defective queries, and rewrites the questions for unambiguous intent. All four metrics now rise together and largely agree. The remaining spread between 89.9 Match and 79.0 SPINACH~F1 is partial-credit granularity, not case-level disagreement.

The progression supports two comparisons. Ontology: on QuAD~3.1, the \texttt{dblpx} wrapper adds +15.0 Match and +25.3 Exact over the native vocabulary (native vs.\ wrapper), with the same model, prompt, and data. The wrapper margin widens on the repaired benchmark (+9.9 to +15.0 Match): with question ambiguity removed, correct parses that previously mismatched the reference are scored as matches. Exact, which forbids supersets, rises more (+25.3 vs.\ +15.0), so Match's superset tolerance does not drive the gain. With the native ontology the model receives a fixed snapshot of the DBLP ontology and the rewriter is inactive. Benchmark: with the \texttt{dblpx} wrapper fixed, the update from 2.0 to 3.1 moves SPINACH~F1 from 17.5 to 79.0. We attribute most of the swing to the deterministic references, with question rewriting and URI pinning contributing the rest. The repair stages were not separately ablated.

\begin{table}[t]
\centering
\caption{DBLP progression from QuAD~2.0 with the native ontology to QuAD~3.1 with the \texttt{dblpx} wrapper. Qwen3.6-27B, procedural prompt, $t{=}0.0$, 1,000 cases, all values \%. The 2.0-wrapper row varies run to run under re-executed references (52.1--52.9 Match over five runs; one shown).}
\label{tab:progression}
\footnotesize
\setlength{\tabcolsep}{3.5pt}
\begin{tabular}{llrrrr}
\toprule
QuAD & Ontology & Match & Exact & QuAD F1 & SPINACH F1 \\
\midrule
2.0 & native & 43.0 & 10.7 & 23.2 & 23.0 \\
2.0 & wrapper & 52.9 & 11.3 & 17.6 & 17.5 \\
2.1 & wrapper & 57.6 & 20.1 & 16.7 & 17.3 \\
3.1 & native & 74.9 & 56.0 & 63.8 & 62.3 \\
3.1 & wrapper & \textbf{89.9} & \textbf{81.3} & \textbf{80.7} & \textbf{79.0} \\
\bottomrule
\end{tabular}
\end{table}

\subsection{Cross-Domain Generality}

Table~\ref{tab:crossdomain} shows results across three domains of different evidential weight: 1,000 cases on DBLP, 100 on SemOpenAlex, and 21 on neuroimaging metadata, the last inherited from the original NLKGQ paper's purpose-built ontology. The same architecture scores 89.9\% Match (reference containment) on the largest and is exact or near-exact on the two smaller sets. Where published baselines exist, we compare on the baseline's own metric: on SemOpenAlex our 98/100 compares against Bartels et al.'s 86/100 (123B model, few-shot with entity mappings)~\citep{bartels2025sparqltrans} on the identical test set. On DBLP our SPINACH~F1 of 79.0\% compares against GRASP's 51.0\% (GPT-4.1, multi-call exploration), with a caveat: GRASP was measured on the original, template-generated DBLP-QuAD~\citep{banerjee2023dblpquad}, while our number is on QuAD~3.1, which removes the entity-resolution and reference-quality confounds documented above. On 2.0 itself our SPINACH~F1 is 17.5 (Table~\ref{tab:progression}). The comparison shows what query generation achieves once the confounds are removed, not a same-benchmark ranking; we expect GRASP and other systems also to score higher on QuAD~3.1, which the released benchmark makes testable.

Auditing the SemOpenAlex test set surfaced two issues. One question appears twice (identical ID, text, and query); we retain the duplicate to preserve the denominator. Seven questions use ``papers that X and Y published'' with a UNION reference query (meaning: all papers by either author), although the natural reading is co-authorship. Our two misses generate the conjunctive reading; disambiguated variants pass. Benchmark ambiguity, not system error, sets the ceiling.

\begin{table}[t]
\centering
\caption{Cross-domain results (Qwen3.6-27B, $t{=}0.0$; DBLP row is the progression run, Table~\ref{tab:progression}). DBLP uses the \texttt{dblpx} wrapper, SemOpenAlex the \texttt{oax} wrapper. The neuroimaging result is the purpose-built-ontology configuration of the original NLKGQ paper~\citep{fitch2026nlkgq}, also with Qwen3.6-27B. Baseline comparisons use the baseline's own metric. The DBLP baseline (GRASP) was measured on 50 samples of the original DBLP-QuAD (see text). F1$_S$ is SPINACH F1, the DBLP baseline's metric.}
\label{tab:crossdomain}
\footnotesize
\setlength{\tabcolsep}{3.5pt}
\begin{tabular}{lrrrrl}
\toprule
Domain & Cases & Match\% & Exact\% & F1$_S$ & Baseline \\
\midrule
DBLP (QuAD 3.1) & 1,000 & 89.9 & 81.3 & 79.0 & 51.0 F1 \\
SemOpenAlex & 100 & 98.0 & 98.0 & --- & 86/100 \\
Neuroimaging & 21 & 100.0 & 100.0 & --- & --- \\
\bottomrule
\end{tabular}
\end{table}

\subsection{Model Comparison: The Concept Layer Outweighs Scale}

Table~\ref{tab:models} compares ten models under identical conditions (same prompt, ontology, temperature). The most consequential result is convergence at the top: two size-matched dense models from independent families reach 90\% Match, Gemma-4-31B (90.3\%) and Qwen3.6-27B-FP8 (90.0\%). Gemma leads on Match only; Qwen3.6-27B is ahead on Exact (81.4\% vs.\ 75.1\%) and both F1 metrics, so Gemma answers more questions correctly but more often with a superset of the reference rows rather than the exact result set. That independent LLM families converge at matched size shows the semantics in the context window matter more than the choice of model family. Mistral-Small-24B, a third family, reaches 80.6\% and beats Llama3.3-70B (74.3\%) with a third of the parameters. Scaling from Qwen3-8B (50.6\% Exact) to Llama3.3-70B (59.4\%), a 9$\times$ cross-family increase in parameters, gains 8.8 Exact points; within the Qwen family, 8B to 27B gains 30.8 Exact points with 3.4$\times$ parameters, confirming that both scale and training matter. Swapping the native vocabulary for the \texttt{dblpx} wrapper on the 27B model gains 25.3 Exact points (56.0$\to$81.3, Table~\ref{tab:progression}) on top of what scale already provides. The progression run and this sweep differ by 0.1 on Exact and both F1 metrics (81.3 vs.\ 81.4 Exact), within run-to-run variation. FP8 quantization matches full precision. The two general-purpose MoE models show high failure rates (25.3\% and 41.5\%), but the code-specialized MoE (Qwen3-Coder-30B) fails only 4.0\% of cases, so sparse activation alone does not explain the failures. Code-focused training appears to compensate. The neuroimaging results follow the general-purpose pattern (35B MoE 57\% vs.\ dense 27B 100\%, summarized above).

\begin{table}[t]
\centering
\caption{Model comparison on QuAD 3.1 + wrapper ($t{=}0.0$, procedural prompt, 1,000 cases). F1$_Q$ is QuAD set-of-values F1. F1$_S$ is SPINACH row-assignment F1 as used by GRASP. All metrics in percent. Fail is the share of cases with no executable query or query error. *partial runs: Qwen3-14B completed 593 and Qwen3.6-35B 730 of 1,000 cases. Their percentages are over completed cases. Frequent query timeouts drive their Fail\%. Type suffixes: /q FP8-quantized, /code code-specialized.}
\label{tab:models}
\footnotesize
\setlength{\tabcolsep}{2pt}
\begin{tabular}{llrrrrr}
\toprule
Model & Type & Match & Exact & F1$_Q$ & F1$_S$ & Fail \\
\midrule
Gemma-4-31B & dense & \textbf{90.3} & 75.1 & 79.4 & 76.9 & 3.5 \\
Qwen3.6-27B-FP8 & dense/q & 90.0 & \textbf{81.4} & \textbf{81.3} & \textbf{79.4} & 2.5 \\
Qwen3.6-27B & dense & 89.9 & \textbf{81.4} & 80.8 & 79.1 & 3.4 \\
Qwen3-Coder-30B & MoE/code & 81.1 & 62.3 & 69.6 & 67.9 & 4.0 \\
Mistral-Small-24B & dense & 80.6 & 65.2 & 68.0 & 64.3 & 4.7 \\
Llama3.3-70B & dense & 74.3 & 59.4 & 63.6 & 61.7 & 5.0 \\
Qwen3-8B & dense & 66.2 & 50.6 & 53.5 & 51.3 & 13.2 \\
Qwen3.6-35B-FP8 & MoE/q & 64.9 & 52.6 & 55.1 & 52.7 & 25.3 \\
Qwen3.6-35B* & MoE & 50.4 & 40.3 & 42.4 & 40.3 & 41.5 \\
Qwen3-14B* & dense & 24.8 & 19.1 & 19.3 & 18.4 & 68.6 \\
\bottomrule
\end{tabular}
\end{table}

\subsection{Error Analysis}

All error analysis uses the progression run (899 Match, Table~\ref{tab:progression} last row). The main failure modes are data coverage and federation scope, not vocabulary confusion. Of the 101 non-matching cases: 24 involve citation queries, where the reference queries return OMID-side entities directly while the wrapper resolves citation endpoints to DBLP publication URIs, so rows diverge wherever the OMID-to-DBLP mapping is incomplete (the 71\% coverage bound); 22 require federated SERVICE calls to DBpedia, Wikidata, or FactGrid, which the \texttt{dblpx} wrapper does not cover (the other 6 federated cases match: their SERVICE blocks need only well-known vocabularies such as FOAF). Of the rest, 18 structure an aggregation differently from the reference; 5 are aggregation queries that failed to execute or timed out; 3 are ontology introspection queries (``what classes exist''); and the remaining 29 are heterogeneous individual errors (venue information dumps, type and label listings, and similar) with no shared mechanism. The run's 30 failed cases (no executable query, or query error) all lie within these buckets: 10 federated, 11 heterogeneous, 5 aggregation, 3 citation, 1 introspection. The sweep run of the same configuration (Table~\ref{tab:models}) fails 34, within run-to-run variation.
\section{Discussion}

\textbf{The concept layer as the dominant factor.}
Our results suggest a hierarchy for LLM query generation: formal concept transfer $>$ model architecture $>$ model scale $>$ prompt engineering. The \texttt{dblpx} wrapper's +25.3 points is additive to model scale on the tested model: the 27B model with the native ontology scores 56.0\%, so scale and wrapper together deliver 81.3\%. Prompt wording sits at the bottom: three system prompt variants tied once the vocabulary was right (summarized in the prior-results subsection), and the concept smearing results below show why prose additions cannot be engineered incrementally. Fine-tuning ranks no higher. An exploratory attempt to train the ontology into an 8B model was not promising: the model lacks the domain vocabulary, which the context window supplies, whereas fine-tuning injects new facts slowly and with more hallucination, and erodes skills outside the tuning distribution~\citep{gekhman2024finetuning,ovadia2024finetuning,kandpal2023longtail,kotha2024forgetting}. \citet{mecharnia2025finetuned} find that fine-tuning succeeds on readable KGs (DBpedia, 61\%) but fails on opaque ones (Wikidata, 13\%). Fine-tuning effort is better spent on general SPARQL-writing skill than on any one domain's vocabulary.

\textbf{Implications beyond SPARQL.}
The principle of maximizing the formal semantic precision of the tokens in the LLM's context should apply to LLM-based applications that generate structured output against a data schema. SQL generation, API call generation, and code generation against typed interfaces all face the same challenge: the LLM must map natural language intent to formal constructs, and the quality of the formal vocabulary it sees determines how well it can do so. The SQL case is measured, not speculative: the neuroimaging experiments of the original NLKGQ paper generated the relational schema from the same ontology, and carrying the ontology's annotations into SQL column comments lifted SQL accuracy from 43\% to 57\%. The concept-transfer levers act across query formalisms; wrapper ontologies are one implementation.

\textbf{Concept smearing: why prompt refinement becomes whack-a-mole.}
At temperature 0.0, LLM SPARQL generation is nearly deterministic, yet any change in prompt tokens shifts many queries it did not target. Large case sets expose this rather than cause it: a ``match names exactly'' instruction fixes its target cases but breaks unrelated queries where the LLM now applies exact matching to DOIs, venue names, and diacritics. In our experiments, rider additions approach zero sum as the rider matures: once the easy corrections are in, remaining errors can only be reached by prose that disturbs cases already correct. Refinement asymptotically approaches whack-a-mole. Our interpretation: natural language tokens are semantically ``wide,'' activating associations across the LLM's full vocabulary and shifting generation probabilities globally. Formal vocabulary tokens are ``narrow'': they participate in specific patterns without broad activation. This is why encoding domain knowledge in formal ontology structure outperforms encoding it in English instructions, and why the rider is reserved for stable conventions rather than incremental repair.

\textbf{Benchmark quality.}
The 61.5-point SPINACH~F1 gap between QuAD~2.0 and QuAD~3.1 (17.5\% vs.\ 79.0\%) demonstrates that benchmark defects drive measured performance.

\textbf{Limitations.}
The wrapper-vs-native ablation runs on a single model (Qwen3.6-27B). The ten-model sweep covers four families but only in the \texttt{dblpx} wrapper configuration. We have not tested proprietary models. Every call carries the complete domain ontology in context. Vocabularies that outgrow the context window would need selection or reduction, which we do not address. Match accepts supersets of the reference rows (Exact, 81.3\%, is the stricter figure), and the QuAD~3.1 repairs and the evaluated system come from the same authors. The per-case logs and pre-repair margin above are the available checks. Wrapper ontology design is manual. Entity resolution is deliberately out of scope: 266 QuAD~3.1 questions carry bracketed URIs their originals lacked, and unbracketed entity mentions remain unsolved. Inherited data limits: only 71\% of DBLP publications carry citation identifiers, and DBLP's maintainers note its ``BibTeX-inherited classification might no longer be a best fit''~\citep{ackermann2024dblpkg}, making some publication-type questions ill-posed.

\section{Conclusion}

Concepts are more effectively passed to LLMs through formal mechanisms than through informal description: OWL ontologies transfer a database schema's concepts as semantically precise tokens that map directly to correct query patterns. Where a native vocabulary is suboptimal, wrapper ontologies substitute precise tokens and a runtime rewriter preserves compatibility, extending the reusable NLKGQ framework from purpose-built domains to existing databases and federations. This layer of controlled semantics contributes more to accuracy than model scaling or prompt revision, tested across three domains and four model families: 89.9\% Match on 1,000 DBLP questions, 98\% on SemOpenAlex, 100\% on neuroimaging metadata. The system prompt tunes query generation; the ontology and rider tune the concepts.

\section*{GenAI Usage Disclosure}

Generative AI tools assisted with editing, \LaTeX{} formatting, coding, and data analysis. A frontier model rewrote the QuAD~3.1 questions; the system under study uses only locally deployed LLMs. All scientific content, experimental design, analysis, and conclusions are the work of the author.

\bibliography{references}

\end{document}